\documentclass[preprint,12pt]{elsarticle}

\usepackage{amssymb}
\usepackage{amsmath,amssymb,amsfonts}
\usepackage{algorithmic}
\usepackage[utf8]{inputenc}
\usepackage{graphicx}
\usepackage{textcomp}
\usepackage{xcolor, colortbl}
\usepackage{tabularx, tabularray}
\usepackage{booktabs}
\usepackage{threeparttable}
\usepackage{graphicx}
\usepackage{enumitem}
\usepackage{enumerate}
\usepackage{array}
\usepackage[T1]{fontenc}
\newcolumntype{P}[1]{>{\centering\arraybackslash}p{#1}}
\newcolumntype{C}[1]{>{\centering\arraybackslash}m{#1}}
\newcolumntype{L}[1]{>{\raggedleft\arraybackslash}p{#1}}
\usepackage{underscore}
\usepackage{adjustbox}
\usepackage{soul}
\usepackage{rotating}
\usepackage{url}
\usepackage{natbib}
\usepackage{afterpage}

\makeatletter
\def\ps@pprintTitle{%
 \let\@oddhead\@empty
 \let\@evenhead\@empty
 \def\@oddfoot{}%
 \let\@evenfoot\@oddfoot}
 \makeatother

\begin{document}
\sloppy

\begin{frontmatter}



\title{Addressing Compiler Errors:\\Stack Overflow or Large Language Models?}


\author[inst1]{Patricia Widjojo}
\author[inst1]{Christoph Treude}
\affiliation[inst1]{organization={University of Melbourne}, 
            city={Parkville},
            postcode={3010}, 
            state={Victoria},
            country={Australia}}


\begin{abstract}
Compiler error messages serve as an initial resource for programmers dealing with compilation errors. However, previous studies indicate that they often lack sufficient targeted information to resolve code issues. Consequently, programmers typically rely on their own research to fix errors. Historically, Stack Overflow has been the primary resource for such information, but recent advances in large language models offer alternatives. This study systematically examines 100 compiler error messages from three sources to determine the most effective approach for programmers encountering compiler errors. Factors considered include Stack Overflow search methods and the impact of model version and prompt phrasing when using large language models. The results reveal that GPT-4 outperforms Stack Overflow in explaining compiler error messages, the effectiveness of adding code snippets to Stack Overflow searches depends on the search method, and results for Stack Overflow differ significantly between Google and StackExchange API searches. Furthermore, GPT-4 surpasses GPT-3.5, with ``How to fix'' prompts yielding superior outcomes to ``What does this error mean'' prompts. These results offer valuable guidance for programmers seeking assistance with compiler error messages, underscoring the transformative potential of advanced large language models like GPT-4 in debugging and opening new avenues of exploration for researchers in AI-assisted programming.
\end{abstract}


\end{frontmatter}


\pagebreak
\section{Introduction}
\label{sec:introduction}

Compiler error messages, designed to guide error resolution, have previously been described as ``difficult to resolve''~\cite{t1}, ``not very helpful''~\cite{t2}, and even ``useless''~\cite{t4}. In fact, a previous study conducted on software engineering students using eye tracking revealed that participants spent between 13 and 25\% of their total task time reading error messages~\cite{t5}, which may suggest inadequacy~\cite{t3} of standard compiler error messages. As a result, developers from novices to experts often rely on external sources during the debugging process~\cite{r1}.

In particular, the Q\&A website Stack Overflow (SO) has often been mentioned as the go-to for programmers searching for coding help including on “why something is failing” ~\cite{r3}, explanations for exceptions ~\cite{r2}, and to repair bugs ~\cite{r7}. Where official documentation is inadequate, Stack Overflow has “often become a substitute for [it]” ~\cite{r8}, and anecdotally reported as having “replaced web search and forums as [users’] primary resource for programming problems” ~\cite{r9}.

However, there exist challenges when searching for Stack Overflow posts relevant to a user’s current compiler error. Error messages are “not standardised for searching related documentation” ~\cite{r1}, the same error may result in different messages and, conversely, the same error message can be produced by “entirely different and distinct errors” ~\cite{r11}. This means that it can be difficult to pinpoint the underlying cause and consequently to create the right Stack Overflow query. Eventually, this may lead to gaps between the user’s intention and the textual query and poor search results ~\cite{r3}. Additionally, there are many cases where multiple people post what is essentially the same question with different words ~\cite{r12}. This, compounded by the sheer number of posts available in Stack Overflow, makes it harder to efficiently locate the information the user is seeking ~\cite{r7, r13}.

More recently, the development and advancement of large language models (LLMs), such as GPT-4~\cite{openai2023gpt4}, have transformed various aspects of technology, including programming and debugging~\cite{sobania2023analysis}. These models have shown the ability to understand natural language, code comprehension, and provide relevant context-sensitive information~\cite{biswas2023role}. As a result, they offer a promising alternative to traditional resources such as SO to address compiler error messages, despite their inability to provide references for their claims and occasional hallucinations~\cite{c1}, which led to ChatGPT's ban on SO~\cite{c2}. Using the knowledge embedded in these AI models, programmers can potentially access more targeted and precise solutions to their debugging challenges. Continuous improvement and expansion of LLMs could potentially revolutionise the way programmers approach and resolve compiler errors.

Our study aims to guide programmers in effectively addressing compiler error messages using two popular tools available right now, SO and LLMs. This paper systematically compares the use of SO and LLMs. We analyse a total of 100 compiler error messages and their context from three sources, aiming to answer key questions such as the optimal search method on SO and the impact of model version and question phrasing when using LLMs. With our first research question, we ask:

\begin{description}
\item[RQ1] How effective are Stack Overflow and large language models at explaining compiler errors?
\end{description}

We find that the LLMs considered in this study, GPT-3.5 and GPT-4, consistently outperform SO in providing explanations for compiler error messages. Additionally, we identify notable performance differences depending on factors such as the inclusion of code snippets in the search rather than just the error message, the method used to access SO content (Google or the StackExchange (SE) API), and the specific version of the LLM employed. To gain a deeper understanding of these aspects and their implications, we conducted a detailed investigation in RQs 2 and 3, examining potential strategies for optimising the debugging process using both SO and LLMs:

\begin{description}
\item[RQ2] To what extent does the query configuration influence Stack Overflow's ability to explain compiler errors?
\end{description}
When seeking assistance with compiler errors on SO, programmers face several decisions: whether to access SO through its own services or a general-purpose search engine like Google, whether to include the offending code in addition to the error message (and if so, how much code to include and whether to remove identifier names that are unlikely to be matched on SO), and whether to focus on accepted or highly voted answers. To answer our second research question, we systematically investigated the impact of these alternatives by designing 72 search strategies as unique combinations of these choices and evaluating their results.

Our findings indicate that when searching SO on Google, omitting code snippets yielded the highest percentage of relevant first answers. In contrast, direct SO searches benefited from the inclusion of code snippets, resulting in increased relevance of results, albeit with the trade-off of fewer results being returned. To measure the similarity of up to the first 10 links returned from a query, we calculate Rank-Biased Overlap (RBO) values~\cite{rbo}. Our analysis shows that searches using Google and the StackExchange (SE) API return entirely different results.

The type of code snippets added to the query significantly affected the similarity of the returned results. By simply removing or replacing the identifiers within the code snippets with ``x'', the RBO with the original query generally decreased to below 0.5. Interestingly, filtering for posts with positive and/or accepted results did not have a notable impact on the overall effectiveness of SO queries. In fact, filtering for results with an accepted answer often led to less relevant first answers.

\begin{description}
\item[RQ3] To what extent does the query configuration influence the ability of large language models to explain compiler errors?
\end{description}

Programmers also face similar decisions when turning to LLMs for help explaining compiler errors: Does the choice of LLM version (e.g., GPT-3.5 vs.~GPT-4) matter, particularly considering the current cost of GPT-4 and initial evidence from other domains suggesting that ChatGPT-4 does not necessarily outperform ChatGPT-3.5 in specific tasks ~\cite{gene}? Do the LLM explanations improve if the offending code is included in the prompt? And what is the most effective way to communicate with the LLM? To address our third research question, we systematically investigate the impact of these alternatives by designing eight prompting strategies as unique combinations of these choices and evaluating their results.

Our findings reveal that GPT-4 outperforms GPT-3.5. When provided with the offending code snippet and the corresponding error message, GPT-4 successfully explains all 100 errors in our dataset, regardless of whether it is asked what the error means or how to fix it. This represents a significant improvement over GPT-3.5, which produced suitable answers in 87\% of the cases for the prompt ``What does the error mean?'' and 91\% for ``How can I fix the error?''. The trend of the ``How can I fix the error?'' prompt leading to more useful results also applies when the LLMs are prompted with only the error message (i.e., without the offending code). GPT-4 produced a suitable response in 84\% of the cases, while GPT-3.5 did so in 75\%. This is both higher than 82\% and 72\% received from their respective ``What'' queries.

These results demonstrate that within a short time span (ChatGPT/GPT-3.5 was released on November 30, 2022, and GPT-4 on March 14, 2023), the models' capabilities in explaining compiler error messages have improved significantly. In particular, the phrasing of the prompt becomes less critical as long as the prompt includes both the offending code and the error message.

In summary, in this paper, we:

\begin{itemize}
\item Conduct a systematic comparison between SO and LLMs (GPT-3.5 and GPT-4) to explain compiler error messages, demonstrating that GPT-4 outperforms both GPT-3.5 and SO in explaining compiler errors, with the inclusion of offending code significantly improving results.
\item Investigate the impact of query configuration on SO's ability to provide relevant answers, considering 72 search strategies and their unique combinations.
\item Assess the influence of prompt configuration on the ability of LLMs to explain compiler errors using eight prompting strategies.
\item Discuss the practical implications of our work for both programmers and researchers.
\end{itemize}

\section{Related Work}
\label{sec:related work}

Programmers frequently conduct search sessions while coding, reflecting their reliance on external resources for assistance during development~\cite{r3}. They tend to prefer general-purpose search engines, particularly Google, over code-specific search engines~\cite{r2, r4, r5, r6}. However, search engines often serve as tools that lead users to other sites, such as SO, CSDN, or ZhiHu, where answers are primarily located~\cite{r2}.

\subsection{Stack Overflow as a Primary Resource}

SO has become an essential resource for programming help, with a large number of questions posted on various topics~\cite{r3}. It often serves as a substitute when official documentation is inadequate, providing solutions to common programming problems~\cite{r8}. However, despite its popularity, finding relevant SO posts can be challenging due to non-standardised error messages~\cite{r1, r11} and the sheer volume of posts, along with duplicated questions~\cite{r7, r12, r13}.

Developers often paste error logs directly into search boxes when looking for answers~\cite{r13}, which may not be the most effective approach, as code queries are more complex and often reformulated or edited~\cite{r3, r5}. They usually rely on Google to find software resources on SO~\cite{r10, r14} and focus on accepted answers when looking at SO posts~\cite{r6, r15, r16}, even though these are not always the most voted-for answers.

Improving search results for error messages has been studied by various researchers, such as Hora, who conducted an empirical study of developer search queries~\cite{r10}, Monperrus and Maia, who focused on JavaScript code snippets in queries~\cite{r17}, Barzilay et al., who explored the types of questions asked and answered on SO~\cite{r8}, Xia et al., who observed developers' use of search engines for code and error explanations~\cite{r2}, and Li et al., who showed that both novice and expert programmers use SO for debugging~\cite{r6}. Our research seeks to identify effective query types for SO, with the goal of reducing the need for reformulation and improving the overall coding experience.

Enhancing compiler error messages has been a focus of previous research, including studies on plugins that add summarised information from SO~\cite{t6, r21} and approaches that improve compiler error messages specifically for Java without using SO~\cite{r22, r23, r24}. Our study complements these findings by testing different factors and assessing the relevance of the results to the original compiler error.

\subsection{Compiler Error Help from Large Language Models}

Artificial intelligence and machine learning have been applied to locate and repair bugs in source code with promising results. DeepFix~\cite{gupta2017deepfix}, a multi-layered sequence-to-sequence neural network with attention, partially fixed almost half of a set of programming tasks by predicting erroneous program locations and providing correct statements. Similar results were achieved with a reinforcement learning approach~\cite{gupta2019deep}. Santos et al.~\cite{santos2018syntax} used n-gram and LSTM language models to locate syntax errors and suggest fixes, achieving similar results. More recently, SYNFIX~\cite{ahmed2021synfix}, which uses unsupervised pre-training and multi-label classification, outperformed previous methods such as DeepFix and Santos et al.

Despite these automatic error-fixing approaches, there has been little prior research on using LLMs to explain compiler error messages. In a recent study based on Codex~\cite{chen2021evaluating}, Leinonen et al.~\cite{leinonen2022using} found that the model produced explanations of error messages that were ``quite comprehensible'', but correct in only about half of all cases. Codex's fix suggestions were also deemed correct in a similar number of cases. To the best of our knowledge, the potential of GPT-4 for explaining compiler error messages has not yet been studied.

\section{Methodology}
\label{sec:methodology}

\subsection{Source Code Collection}

To cover the varying use cases of SO and LLMs when addressing different compiler errors, we gathered Java source code with compiler errors from multiple sources, totaling 100 pieces. First, we randomly obtained 55 pieces of code from the Blackbox dataset~\cite{t22}. The Blackbox dataset comprises activity data from BlueJ IDE users~\cite{r22} who consented to have their activity recorded for research purposes. Using these pieces of code, we consider real-world instances where developers encountered compiler error messages during their coding projects. Additionally, we expanded the range of errors by creating 18 sets of code with unique error types not yet encountered in the BlackBox data. These pieces of code were designed to fail due to frequently occurring Java compiler errors, as documented in~\cite{r23, r24}, made loosely based on~\cite{t23,new1}, and had not been encountered in the first batch of snippets from Blackbox (code number 1-35).  Lastly, 27 pieces of source code were sourced from a separate participant study conducted as part of another project by the first author. This project also focused on compiler errors, and the code was collected when three participants with diverse backgrounds and levels of Java experience completed exercises from code-exercises.com~\cite{t24}. In the results and discussion sections, we will refer to these three sources of Java code as \textsc{blackbox}, \textsc{custom}, and \textsc{user\_study}, respectively.  

\begin{figure}[h]
\centering
\includegraphics[width=\linewidth]{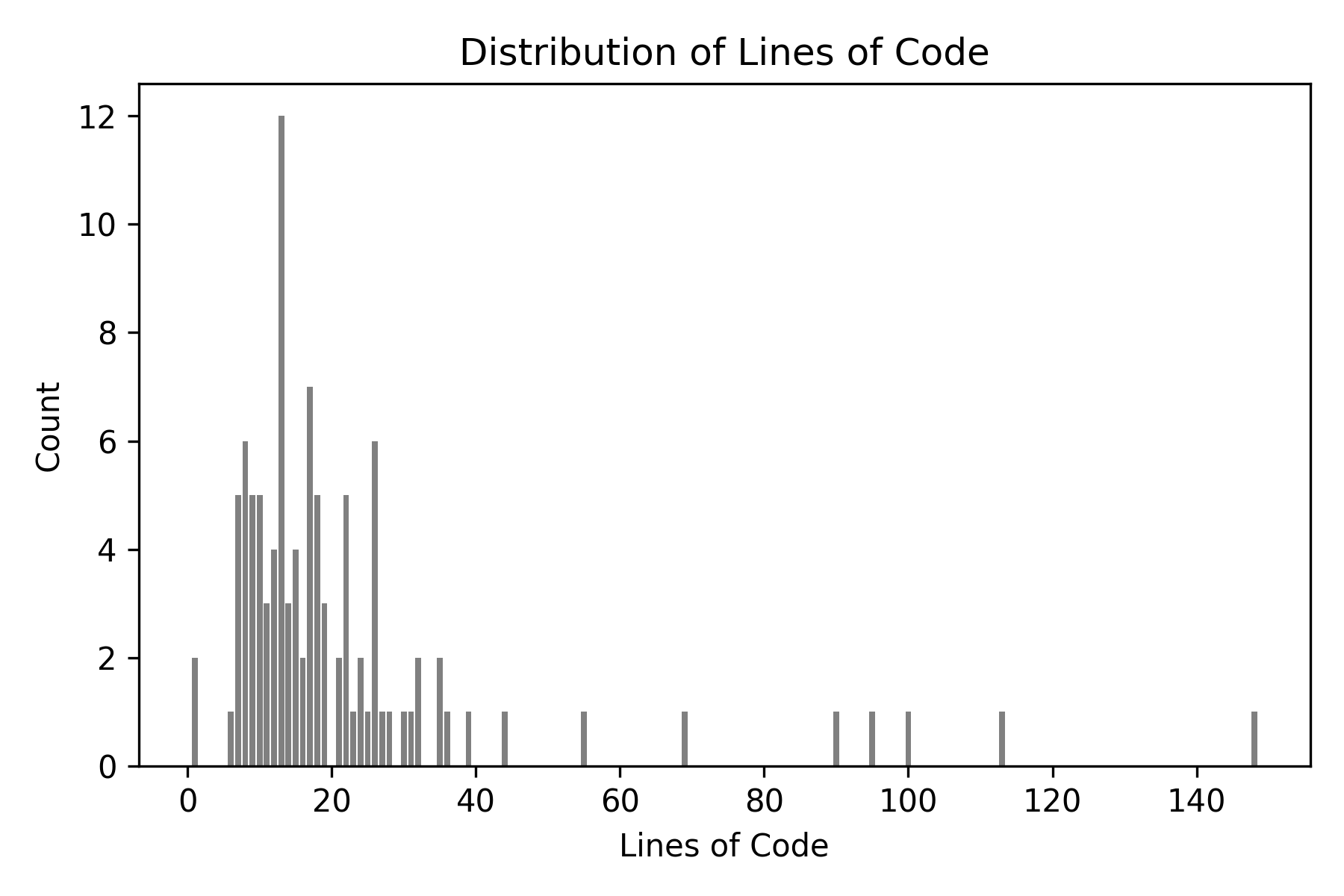}
\caption{Lines of Code (LOC) distribution of code snippets}
\label{fig: loc}
\end{figure}

We sourced code snippets from three origins to increase confidence in generalisability and to ensure our dataset accurately represents a variety of real-world coding errors. As the codes were retrieved from singular files as opposed to a project with multiple files, the length of each is relatively short with a median Lines of Code (LOC) of 16. The complete LOC distribution is shown in Figure \ref{fig: loc}.

\subsection{Evaluation Methodology}

To analyse the impact of various factors on the overall effectiveness of SO and LLM queries, we evaluate the results obtained under different configurations. We define effectiveness as the degree of assistance provided by a result in resolving the first compiler error encountered in a Java program.

For SO queries, we classified the resulting initial posts into four distinct relevance categories, as detailed in Table~\ref{table: ih table}. This rating scale was proposed and used by Mahajan et al.~\cite{r20} in measuring the relevance of SO posts, and was based on~\cite{o2,o20,o15}. We refer to this categorisation as the IHMU-category.

\renewcommand{\arraystretch}{1.1}
\begin{table}[ht]
    \centering
    \caption{IHMU-category relevancy classification}
    \label{table: ih table}
    \footnotesize
\begin{tabular}{m{3.6cm}m{9.2cm}}
\bottomrule
    \hline
\rowcolor{gray!20} 
\textbf{Classification} & \textbf{Definition}  \\
\hline
\textbf{Instrumental (I)}   & Post relates to the underlying cause of the first compiler error perfectly and provides a complete fix \\
 \hline
 \textbf{Helpful (H)} & Post is generally informative about the compiler error but may not specifically target the underlying cause or provide an effective fix  \\
 \hline
\textbf{Misleading (M)} &  Post is irrelevant to the compiler error. \\
 \hline
\textbf{Unavailable (U)} & No results were returned from the query   \\
 \bottomrule
  \hline
\end{tabular}
    
\end{table}

During the classification of the results, we considered the entire content of the page to mimic real-world usage, where developers are more likely to focus on the text of the post rather than just the titles~\cite{o0,r6}. The first author conducted this manual annotation. A cross-check with the second author annotating a subsection (10\%) of the results was also performed. The Cohen's Kappa~\cite{kappa} coefficient was then calculated to measure inter-rater reliability, ensuring that the overall annotation bias is acceptable and that the results can be considered valid. We obtained a Cohen's Kappa value of 0.860, indicating an agreement level of ``almost perfect''.

After completing all necessary SO annotations, we followed Mahajan et al.'s evaluation methodology and calculated three metrics: I-Score, IH-Score, and M-Score. Additionally, we introduced an IH/M-Score, which measures the percentage of instrumental and helpful results relative to the number of misleading results. This metric helps evaluate whether an increase in relevant results is accompanied by a trade-off of also increasing misleading results. These metrics and their respective formulas are listed in Table~\ref{table: metric}.

\renewcommand{\arraystretch}{1.8}
\begin{table}[ht]
    \centering
    \caption{Formulas used in measuring result relevancy}
    \label{table: metric}
    \footnotesize
\begin{tabular}{m{3.6cm}m{9.2cm} }
\bottomrule
    \hline
\rowcolor{gray!20} 
\textbf{Classification} & \textbf{Definition}  \\
\hline
 \textbf{I-Score} & $\tfrac{I}{I+H+U+M} \times 100\%  $ \\
 \hline
 \textbf{IH-Score} & $\tfrac{I+H}{I+H+U+M} \times 100\%  $ \\
 \hline
\textbf{M-Score} & $ \tfrac{M}{I+H+U+M} \times 100\% $\\
 \hline
 \textbf{IH/M-Score} & $ \tfrac{I+H}{M} \times 100\% $\\
 \bottomrule
  \hline
\end{tabular}
    
\end{table}

Moreover, we measured the similarity between the results of different SO combinations. To achieve this, we calculated the Rank-Biased Overlap (RBO)~\cite{rbo} of the top 10 links for each combination in the 100 scenarios. An RBO score of 1 indicates a completely ``identical'' list, while a score of 0 indicates a completely ``different'' list~\cite{rbo}. We then computed the median of these scores and present the corresponding heatmap.

The evaluation of the results obtained from LLMs followed a similar methodology to that of SO, with a few adjustments. Since ChatGPT, as a chatbot on top of GPT-3.5 and GPT-4, virtually always returns results that are or at least appear relevant when provided with error messages, we modified the IHMU-category to exclude the ``Unavailable'' category and more narrowly defined the other categories, as shown in Table~\ref{table: gpt ih table}. We did not calculate similarities through RBO between the results obtained from LLMs as part of this study.

\renewcommand{\arraystretch}{1.3}
\begin{table}[ht]
    \centering
    \caption{Adjusted IHM-category relevancy classification}
    \label{table: gpt ih table}
    \footnotesize
\begin{tabular}{m{3.6cm}m{9.2cm}}
\bottomrule
    \hline
\rowcolor{gray!20} 
\textbf{Classification} & \textbf{Definition}  \\
\hline
\textbf{Instrumental (I)}   & Response perfectly targets underlying cause of error and provides a clear action on how to fix \\
 \hline
 \textbf{Helpful (H)} & Response provides general but not exact help. E.g. listing all potential common causes of error along with multiple fixes from which the coder would still have to analyse \\
 \hline
\textbf{Misleading (M)} &  Response does not provide a clear direction on how to fix the issue and/or causes confusion. E.g. irrelevant results, right code snippet however with a completely wrong explanation, returning other common causes without the root cause, and responses that provides fixes to errors in different parts of the code but would not fix the first compiler error. \\
 \bottomrule
  \hline
\end{tabular}
    
\end{table}

\subsection{Stack Overflow Combinations}

\renewcommand{\arraystretch}{1.2}
\begin{table*}
\centering
\caption{Description of the identifiers used for SO via API and SO via Google queries}
\label{table: id table}
\adjustbox{max width=\textwidth}{
\small 
\begin{tabular}{ m{3cm}m{5cm}m{10cm} }
\bottomrule
\hline
\rowcolor{gray!60} 
\multicolumn{3}{c}{\textbf{Query Method}}\\
\hline
\rowcolor{gray!20} 
\textbf{Identifier }& \multicolumn{2}{l}{\textbf{Query Search Method}} \\
\hline
\textsc{SO-API} & \multicolumn{2}{l}{Search request through StackExchange API as a proxy for searching Stack Overflow directly} \\
\hline
\textsc{SO-Google} &  \multicolumn{2}{l}{Search via Google with ``site: stackoverflow'' included in query} \\
\bottomrule
\hline
\rowcolor{gray!60} 
\multicolumn{3}{c}{\textbf{Post Filter}}\\
\hline
\rowcolor{gray!20} 
\textbf{Identifier }& \multicolumn{2}{l}{\textbf{Query Result Filter}} \\
\hline
\textsc{First} &  \multicolumn{2}{l}{Take first result (Stack Overflow post) without any filtering} \\
\hline
\textsc{Positive} &  \multicolumn{2}{l}{Take first result (Stack Overflow post) with a positively rated answer} \\
\hline
\textsc{Accepted} &  \multicolumn{2}{l}{Take first result (Stack Overflow post) with an accepted answer} \\
\bottomrule
\hline
\rowcolor{gray!60} 
\multicolumn{3}{c}{\textbf{Query Content}}\\
\hline
\rowcolor{gray!20} 
\textbf{Identifier} & \textbf{Description}  & \textbf{Example}\\ 
\hline
\begin{tabular}{@{}l@{}}\textsc{Custom/} \\ \textsc{Line}\end{tabular} & custom error message + original error line & non-static method cannot be referenced from a static context System.out.println(reverse(s)); \\
\hline
\begin{tabular}{@{}l@{}}\textsc{Custom/} \\ \textsc{LineBlankId}\end{tabular} & custom error message + error line with identifiers replaced with blanks & non-static method cannot be referenced from a static context ..(()); \\
\hline
\begin{tabular}{@{}l@{}}\textsc{Custom/} \\ \textsc{NoSnippet}\end{tabular}  & custom errror message only  & non-static method cannot be referenced from a static context \\
\hline
\begin{tabular}{@{}l@{}}\textsc{Custom/} \\ \textsc{Parent}\end{tabular}  & custom error message + parent node of error line & non-static method cannot be referenced from a static context [    public static void main(){,         String s = "java interview";,         System.out.println());,     }] \\
\hline
\begin{tabular}{@{}l@{}}\textsc{Custom/} \\ \textsc{ParentBlankId}\end{tabular}  & custom error message + parent node with identifiers replaced with blanks & non-static method cannot be referenced from a static context public static void main(){ String  = "java interview"; ..()); }  \\
\hline
\begin{tabular}{@{}l@{}}\textsc{Custom/} \\ \textsc{ParentXId}\end{tabular}  & custom error message + parent node with identifiers replaced with "x" & non-static method cannot be referenced from a static context public static void main(){ String x = "java interview"; x.x.x()); }   \\
\hline
\begin{tabular}{@{}l@{}}\textsc{Original/} \\ \textsc{Line}\end{tabular}  & original error message + original error line & non-static method reverse(String) cannot be referenced from a static context System.out.println(reverse(s));\\
\hline
\begin{tabular}{@{}l@{}}\textsc{Original/} \\ \textsc{LineBlankId}\end{tabular} & original error message +  error line with identifiers replaced with blanks & non-static method reverse(String) cannot be referenced from a static context ..(()); \\
\hline
\begin{tabular}{@{}l@{}}\textsc{Original/} \\ \textsc{NoSnippet}\end{tabular}  & original error message only & non-static method reverse(String) cannot be referenced from a static context  \\
\hline
\begin{tabular}{@{}l@{}}\textsc{Original/} \\ \textsc{Parent}\end{tabular}  & original error message + parent node of error line & non-static method reverse(String) cannot be referenced from a static context [    public static void main(){,         String s = "java interview";,         System.out.println());,     }] \\
\hline
\begin{tabular}{@{}l@{}}\textsc{Original/} \\ \textsc{ParentBlankId}\end{tabular}  & original error message + parent node with identifiers replaced with blanks & non-static method reverse(String) cannot be referenced from a static context public static void main(){ String  = "java interview"; ..()); }   \\
\hline
\begin{tabular}{@{}l@{}}\textsc{Original/} \\ \textsc{ParentXId}\end{tabular}  & original error message + parent node with identifiers replaced with "x" & non-static method reverse(String) cannot be referenced from a static context public static void main(){ String x = "java interview"; x.x.x()); }    \\
\bottomrule
\hline
\end{tabular}
}
\end{table*}

Given the numerous ways to search SO, we conducted experiments with various query approaches to identify the most effective methods for obtaining relevant results. Each query corresponded to a different combination of query method, post filter, and query content, resulting in a total of $2\times3\times12=72$ unique combinations. These are described in Table~\ref{table: id table}, with additional definitions and clarifications outlined below.

First, to obtain results with ``searching directly through SO'' (\textsc{SO-API}) as a search method, we used the /search functionality of the SE API~\cite{stackexchange}, with the corresponding queries saved under the ``intitle'' argument. We employed this method because of the absence of a full-text search option.

The \textsc{Custom} query content treatment refers to queries in which we replaced code specific identifiers from the original compiler error message with a more generic variable or removed it altogether. This modification was implemented for some common messages to make them more generic. For example, we replaced the variable and method name in the ``variable [variable name] is already defined in method [method name]'' errors, and the class name in the ``class [class name] is public, should be declared in a file named [class name].java'' errors with generic placeholders ``X'' and ``Y''. 

Under query content, the ``parent node'' was obtained by creating a concrete syntax tree of the Java code snippet using the parser library tree-sitter~\cite{tree-sitter} and then saving the parent node of the line where the compiler error occurred. The same parser library was used to identify parts of the code labelled as identifiers, which were used in defining \textsc{LineBlankId, ParentBlankId}, and \textsc{ParentXId}.

We tested each of these 72 combinations on the 100 Java codes as described in the Source Code Collection section, yielding a total of $72\times100=7,200$ query runs. From each run, we saved up to the first 10 SO links returned for further analysis of relevance and similarity. 

To facilitate data collection, we created a Sublime Text 4 plugin. When called, the plugin automatically executes each of the 72 combinations described in Table~\ref{table: id table} and logs the details related to each run. This information included the combination to which it corresponded, the source code on which it was run, the total number of results returned, and up to the first 10 SO links it returned. The total number of results returned for queries run on \textsc{SO-API} was simply counted as the number of items in the JSON file it returned, whereas for Google \textsc{SO-Google} queries, we scraped the resulting page to find the div element with the id ``result-stats'' from which the relevant value was saved. This represents the part ``About [number] results'' found in Google search results. When errors occurred while scraping the page, we placed a placeholder value of 0 instead. This placeholder value is not expected to affect the results, as the percentage of error occurrence was negligible at only 0.5\%. We used the plugin on each of the 100 Java pieces of code with compiler errors as previously described.

Compiling the first links from all 7,200 query runs yielded a total of 573 unique pairs of source code and the corresponding first link. The relatively lower number of unique pairs was due to several combinations returning the same top link and many combinations, particularly those with long code snippets, not returning any results.

\subsection{Large Language Model Combinations}

\renewcommand{\arraystretch}{1.2}
\begin{table*}
\centering
\caption{Description of the identifiers used for LLM queries}
\label{table: id table gpt}
\adjustbox{max width=\textwidth}{
\small 
\begin{tabular}{ m{3cm}m{5cm}m{10cm} }

\bottomrule
\hline
\rowcolor{gray!60} 
\multicolumn{3}{c}{\textbf{Query Method}}\\
\hline
\rowcolor{gray!20} 
\textbf{Identifier }& \multicolumn{2}{l}{\textbf{Query Search Method}} \\
\hline
\textsc{GPT-3.5} & \multicolumn{2}{l}{Query on https://chat.openai.com/ with ``Default (GPT-3.5)'' model} \\
\hline
\textsc{GPT-4} & \multicolumn{2}{l}{Query on https://chat.openai.com/ with ``GPT-4'' model} \\

\bottomrule
\hline
\rowcolor{gray!60} 
\multicolumn{3}{c}{\textbf{Query Phrasing}}\\
\hline
\rowcolor{gray!20} 
\textbf{Identifier}& \multicolumn{2}{l}{\textbf{Query Phrasing}} \\
\hline
\textsc{What} &  \multicolumn{2}{l}{Query content asked alongside ``What does the error mean?''} \\
\hline
\textsc{How} &  \multicolumn{2}{l}{Query content asked alongside ``How can I fix the error?''} \\

\bottomrule
\hline
\rowcolor{gray!60} 
\multicolumn{3}{c}{\textbf{Query Content}}\\
\hline
\rowcolor{gray!20} 
\textbf{Identifier} & \textbf{Description} & \textbf{Example}\\ 
\hline
\begin{tabular}{@{}l@{}}\textsc{Original/} \\ \textsc{FullSnippet}\end{tabular} & original error message + full Java snippet from which the first error message is thrown. & test.java:5: error: cannot find symbol public class test { public static void main(String[] args) { System.ouch.println("Hello, World!"); } } \\
\hline
\begin{tabular}{@{}l@{}}\textsc{Original/} \\ \textsc{NoSnippet}\end{tabular} & original error only & test.java:5: error: cannot find symbol \\
\bottomrule
\hline
\end{tabular}
}
\end{table*}

To address research questions related to LLMs, we examined the impact of including the source code in the query versus only providing the error message, the influence of prompt phrasing on the relevance of the results, and the differences between using the ``Default (GPT-3.5)'' model and the newer ``GPT-4'' version. We selected the prompts ``What does the error mean?'' and ``How can I fix the error?'' based on experiments with GPT-3.5. Table~\ref{table: id table gpt} enumerates the $2\times2\times2=8$ LLM query combinations.

We collected LLM data by pasting each of the 100 error messages with and without code snippet on the ChatGPT page\footnote{https://chat.openai.com} along with the one of the two prompts. To maintain consistency, only the first responses i.e. no regeneration of the 800 queries were saved and evaluated.

\subsection{Data Availability}

We have made our data and scripts available in our online repository\footnote{https://github.com/patwdj/java-compiler-error-help}. This includes an Excel file containing the 100 code snippets tested along with their corresponding compiler errors, resulting SO links, LLM responses, authors' annotations, and resulting tables. Additionally, the two notebooks used to calculate the median number of results and RBOs, as well as the Sublime plugin created to obtain links based on the 72 combinations related to SO, have been made available.

\section{Results}
\label{sec:results}

In this section, we present and discuss answers to our research questions. An important factor to consider is that \textsc{SO-API} queries frequently returned no results, affecting the results. Out of 100 scenarios, the average number of times SE queries yielded no results was 83.1. When queries without code snippets are excluded, empty results become even more prominent, averaging about 91.1\%. On the contrary, the average empty results for \textsc{SO-Google} queries were significantly lower at 23.7\% in general or 27.8\% when excluding queries without code snippets. 

Table~\ref{table: number of results} presents the median number of results for the 100 compiler errors when using different query combinations. Since the median values for \textsc{SO-API} queries are mostly 0, and LLMs provide one response per query, the discussion about the number of results will mainly focus on \textsc{SO-Google}.

\renewcommand{\arraystretch}{1.2}
\begin{table}[h]
    \centering
    \caption{Median number of results returned}
    \label{table: number of results}
    \footnotesize
    \begin{tabular}{m{9.2cm}L{3.6cm}}
    \bottomrule
    \hline
    \rowcolor{gray!20}
    \textbf{Identifier} &  \textbf{Median} \\ \hline
    \textsc{SO-Google/Custom/NoSnippet} & 36350 \\ \hline
    \textsc{SO-Google/Original/NoSnippet}  & 32400\\ \hline
    \textsc{SO-Google/Custom/LineBlankId} & 2570 \\ \hline
    \textsc{SO-Google/Original/LineBlankId} & 2360\\ \hline
    \textsc{SO-Google/Custom/LineXId} & 996.5\\ \hline
    \textsc{SO-Google/Original/LineXId} & 813.5\\ \hline
    \textsc{SO-Google/Original/Line}  & 471 \\ \hline
    \textsc{SO-Google/Custom/Line} & 372.5 \\ \hline
    \textsc{SO-Google/Custom/ParentBlankId}  & 163.5\\ \hline
    \textsc{SO-Google/Original/ParentBlankId} & 8\\ \hline
    \textsc{SO-Google/Custom/Parent} & 6.5\\ \hline
    \textsc{SO-Google/Original/Parent} & 6 \\ \hline \hline
    \textsc{SO-API/Custom/NoSnippet} & 3.5 \\ \hline
    \textsc{SO-API/Original/NoSnippet}  & 2\\ \hline
    \textsc{SO-API/...} & 0 \\ \bottomrule\hline
    \end{tabular}
\end{table}

\renewcommand{\arraystretch}{1.2}
\begin{table}[h]
    \centering
    \caption{Summary of Results}
    \label{table: results rq1}
    \footnotesize
    \begin{tabular}{m{8.4cm}L{1.2cm}L{1.2cm}L{1.2cm} }
    \bottomrule
    \hline
    \rowcolor{gray!20}
        \textbf{Search (all with original error message)} &  \textbf{IH} & \textbf{M} &  \textbf{U} \\ \hline
        \textsc{SO-API/NoSnippet} & 15 & 40 & 45 \\ \hline
        \textsc{SO-Google/NoSnippet} & 57 & 41 & 2 \\ \hline
        \textsc{SO-API/Parent} & 4 & 0 & 96 \\ \hline
        \textsc{SO-Google/Parent} & 22 & 48 & 30 \\ \hline
        \textsc{GPT-3.5/What/NoSnippet} & 72 & 28 & - \\ \hline
        \textsc{GPT-4/What/NoSnippet} & 82 & 18 & - \\ \hline
        \textsc{GPT-3.5/What/FullSnippet} & 87 & 13 & - \\ \hline
        \textsc{GPT-4/What/FullSnippet} & 100 & 0 & - \\ \bottomrule\hline
    \end{tabular}
\end{table}

More than the sheer number of results, when using SO or LLMs to resolve compiler errors, the relevancy of a post or response to the specific issue is important. We have summarised the values of the IHM(U) categories, as defined in Table~\ref{table: ih table} and~\ref{table: id table gpt}, in Table~\ref{table: results rq1}. This summary only includes a subset of the combinations tested and will be discussed in relation to RQ1. A more detailed breakdown of other \textsc{SO-Google} and \textsc{SO-API} combinations, along with their corresponding metrics, will be presented as part of the RQ2 results, while GPT-3.5 and GPT-4 will be discussed as part of RQ3.

\subsection*{\textbf{[RQ1] }How effective are Stack Overflow and large language models at explaining compiler errors?}

Table~\ref{table: results rq1} shows that SO queries exhibit mixed results in their ability to help explain compiler errors when searched directly through the API and when searched through Google. Without code snippets, searching through Google is more promising, with 57\% IH compared to \textsc{SO-API/NoSnippet}'s 15\%. The high number of misleading compared to the relevant relevant results is also particularly concerning for \textsc{SO-API/NoSnippet}. In contrast, when the parent node of the error line is included in the query, the trend reverses: \textsc{SO-API/Parent} does not return misleading results, compared to over 40\% found by its \textsc{SO-Google} counterpart. However, although \textsc{SO-API/Parent} did not give misleading results, it only returned results in 4 of the 100 scenarios.

In terms of seeking help for Java compiler errors, LLMs proves to be notably more effective than SO. LLMs consistently return a result with a higher relevancy rate than our SO searches. Specifically, when using GPT-4 and providing the full Java code snippet along with the original error message, a relevancy rate of 100\% is achieved. This represents an improvement over the older GPT-3.5 version, which achieved a relevancy rate of 87\% for the same queries. A consistent trend observed in both versions is that adding the full code snippet to the query positively impacts the results. In scenarios with or without code snippets in the query, LLM's results consistently outperform SO's.

\subsection*{\textbf{[RQ2] }To what extent does the query configuration influence Stack Overflow's ability to explain compiler errors?}

\begin{figure}[h]
\centering
  \includegraphics[width = \linewidth]{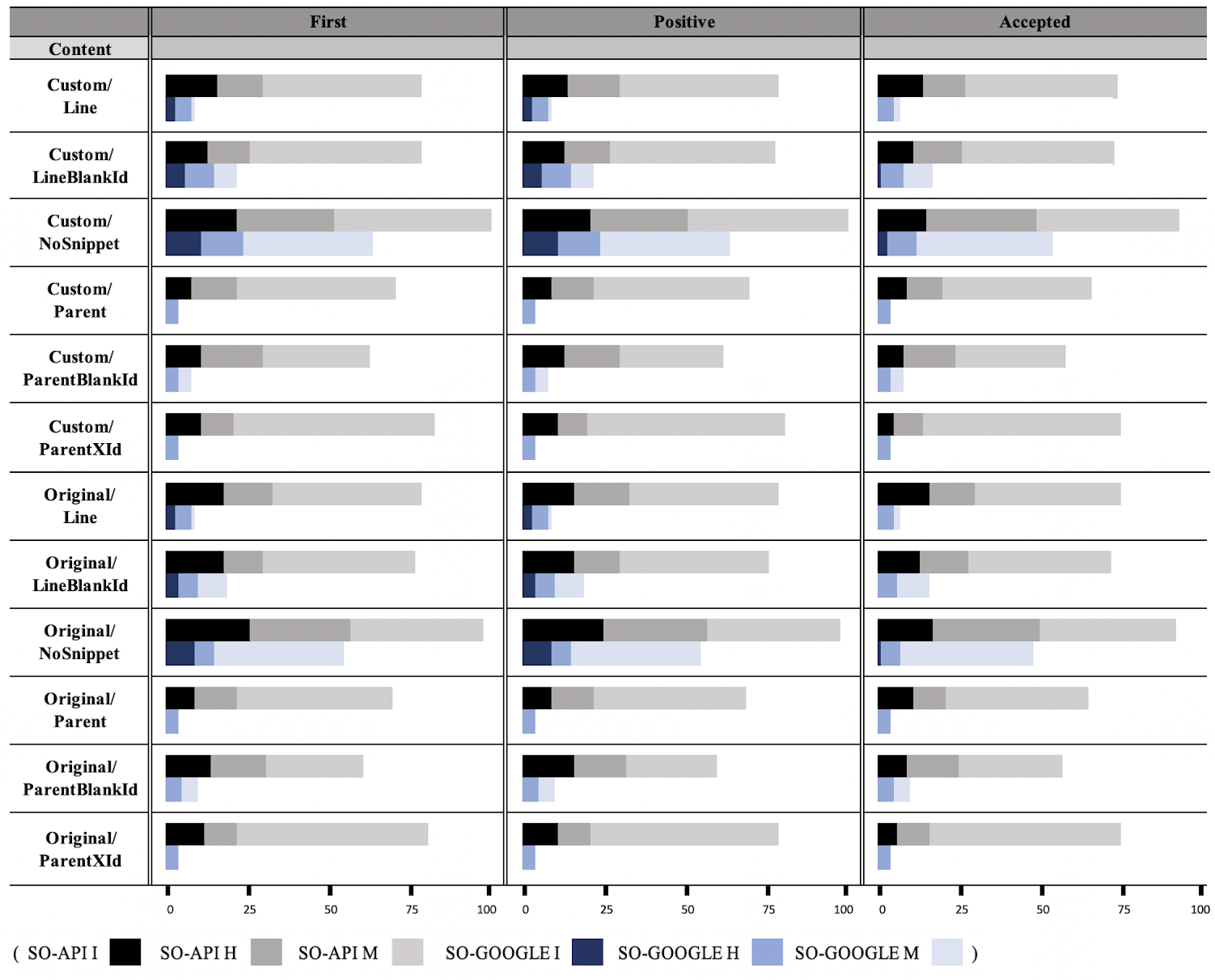}
  \caption{IHMU-category of resulting first links}
  \label{fig: result stacked chart}
  
\end{figure}

We evaluated 72 different SO query combinations. The percentage of results for the different combinations based on their respective IHMU categories is shown in Figure~\ref{fig: result stacked chart}. The corresponding metrics can be found in Table~\ref{table: result scores}.

  \clearpage
\renewcommand{\arraystretch}{1.3}
\begin{sidewaystable}[h]
    \scriptsize
    \caption{Calculated relevancy metrics}
    \label{table: result scores}  
\begin{tabular}
{m{5.6cm}||P{0.6cm}P{0.6cm}P{0.6cm}P{0.85cm}||P{0.6cm}P{0.6cm}P{0.6cm}C{0.85cm}||P{0.6cm}P{0.6cm}P{0.6cm}P{0.85cm}}
\bottomrule
\hline
\rowcolor{gray!60} 
 & \multicolumn{4}{c}{\textbf{First (Scores)}} & \multicolumn{4}{c}{\textbf{Positive (Scores)}} & \multicolumn{4}{c}{\textbf{Accepted (Scores)}} \\ \hline
 \rowcolor{gray!20} 
 & \textbf{I} & \textbf{IH} & \textbf{M} & \textbf{IH/M} & \textbf{I} & \textbf{IH} & \textbf{M} & \textbf{IH/M} & \textbf{I} & \textbf{IH} & \textbf{M} & \textbf{IH/M}  \\ \hline
\textsc{SO-Google/Custom/Line} & 16\% & 30\% & 49\% & 61\% & 14\% & 30\% & 49\% & 61\% & 14\% & 27\% & 47\% & 57\%  \\ 
\textsc{SO-API/Custom/Line} & 3\% & 8\% & 1\% & 800\%  & 3\% & 8\% & 1\% & 800\%  & 0\% & 5\% & 2\% & 250\%  \\ \hline
\textsc{SO-Google/Custom/LineBlankId} & 13\% & 26\% & 53\% & 49\%  & 13\% & 27\% & 51\% & 53\%  & 11\% & 26\% & 47\% & 55\%  \\ 
\textsc{SO-API/Custom/LineBlankId} & 6\% & 15\% & 7\% & 214\%  & 6\% & 15\% & 7\% & 214\%  & 1\% & 8\% & 9\% & 89\% \\ \hline
\textsc{SO-Google/Custom/NoSnippet} & 22\% & 52\% & 48\% & \textbf{108\%}  & 21\% & 51\% & 49\% & \textbf{104\%} & 15\% & 49\% & 44\% & \textbf{111\%}  \\ 
\textsc{SO-API/Custom/NoSnippet} & 11\% & 24\% & 40\% & 60\% & 11\% & 24\% & 40\% & 60\%  & 3\% & 12\% & 42\% & 29\%  \\ \hline
\textsc{SO-Google/Custom/Parent} & 8\% & 22\% & 49\% & 45\%  & 9\% & 22\% & 48\% & 46\%  & 9\% & 20\% & 46\% & 43\%  \\ 
\textsc{SO-API/Custom/Parent} & 0\% & 4\% & 0\% & \textbf{inf}  & 0\% & 4\% & 0\% & \textbf{inf}  & 0\% & 4\% & 0\% & \textbf{inf}  \\ \hline
\textsc{SO-Google/Custom/ParentBlankId} & 11\% & 30\% & 33\% & 91\%  & 13\% & 30\% & 32\% & 94\% & 8\% & 24\% & 34\% & 71\%  \\ 
\textsc{SO-API/Custom/ParentBlankId} & 0\% & 4\% & 4\% & 100\%  & 0\% & 4\% & 4\% & 100\% & 0\%  & 4\% & 4\% & 100\%  \\ \hline
\textsc{SO-Google/Custom/ParentXId} & 11\% & 21\% & 62\% & 34\%  & 11\% & 20\% & 61\% & 33\% & 5\% & 14\% & 61\% & 23\% \\ 
\textsc{SO-API/Custom/ParentXId} & 0\% & 4\% & 0\% & \textbf{inf}  & 0\% & 4\% & 0\% & \textbf{inf} & 0\% & 4\% & 0\% & \textbf{inf} \\ \hline
\hline
\textsc{SO-Google/Original/Line} & 18\% & 33\% & 46\% & 72\% & 16\% & 33\% & 46\% & 72\%  & 16\% & 30\% & 45\% & 67\% \\ 
\textsc{SO-API/Original/Line} & 3\% & 8\% & 1\% & 800\% & 3\% & 8\% & 1\% & 800\%  & 0\% & 5\% & 2\% & 250\% \\ \hline
\textsc{SO-Google/Original/LineBlankId} & 18\% & 30\% & 47\% & 64\% & 16\% & 30\% & 46\% & 65\%  & 13\% & 28\% & 44\% & 64\% \\ 
\textsc{SO-API/Original/LineBlankId} & 4\% & 10\% & 9\% & 111\% & 4\% & 10\% & 9\% & 111\%  & 0\% & 6\% & 10\% & 60\% \\ \hline
\textsc{SO-Google/Original/NoSnippet} & 26\% & 57\% & 41\% & \textbf{139\%}  & 25\% & 57\% & 41\% & \textbf{139\%}  & 17\% & 50\% & 42\% & \textbf{119\%} \\ 
\textsc{SO-API/Original/LineBlankId} & 9\% & 15\% & 40\% & 38\% & 9\% & 15\% & 40\% & 38\%  & 1\% & 7\% & 41\% & 17\% \\ \hline
\textsc{SO-Google/Original/Parent} & 9\% & 22\% & 48\% & 46\%  & 9\% & 22\% & 47\% & 47\%  & 11\% & 21\% & 44\% & 48\%  \\ 
\textsc{SO-API/Original/Parent} & 0\% & 4\% & 0\% & \textbf{inf}  & 0\% & 4\% & 0\% & \textbf{inf} & 0\% & 4\% & 0\% & \textbf{inf}  \\ \hline
\textsc{SO-Google/Original/ParentBlankId} & 14\% & 31\% & 30\% & 103\%  & 16\% & 32\% & 28\% & 114\%  & 9\% & 25\% & 32\% & 78\% \\ 
\textsc{SO-API/Original/ParentBlankId} & 0\% & 5\% & 5\% & 100\% & 0\%  & 5\% & 5\% & 100\% & 0\%  & 5\% & 5\% & 100\%  \\ \hline
\textsc{SO-Google/Original/ParentXId} & 12\% & 22\% & 59\% & 37\% & 11\% & 21\% & 58\% & 36\%  & 6\% & 16\% & 59\% & 27\%  \\ 
\textsc{SO-API/Original/ParentXId} & 0\% & 4\% & 0\% & \textbf{inf}  & 0\% & 4\% & 0\% & \textbf{inf}  & 0\% & 4\% & 0\% & \textbf{inf}  \\ \bottomrule\hline
\end{tabular}
\end{sidewaystable}
\clearpage 

\begin{figure}[h]
\centering
\includegraphics[width=\linewidth]{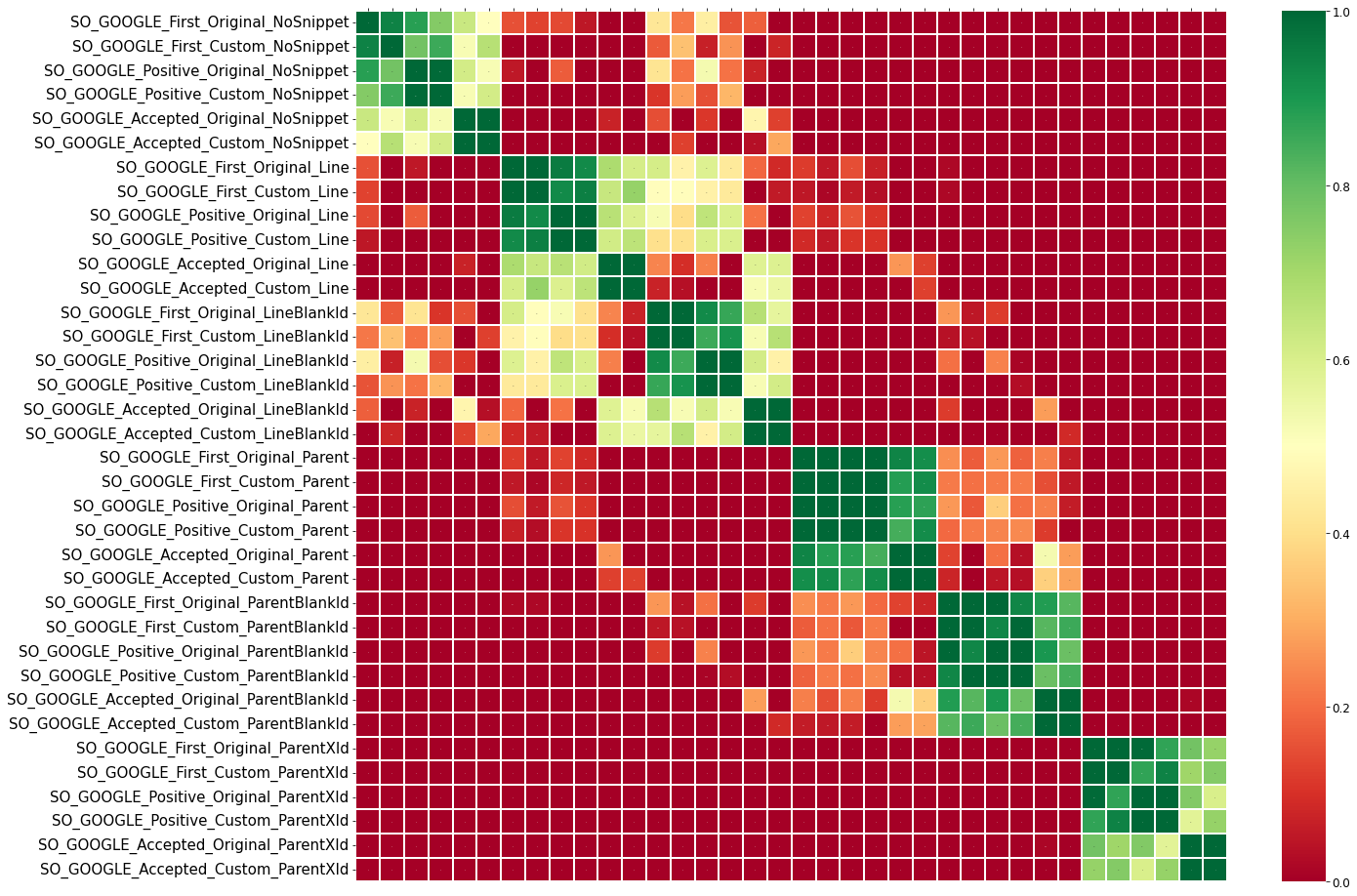}
\caption{RBO values between Google \textsc{SO-Google} queries}
\label{fig: rbog}
\end{figure}

\begin{figure}[h]
\centering
\includegraphics[width=\linewidth]{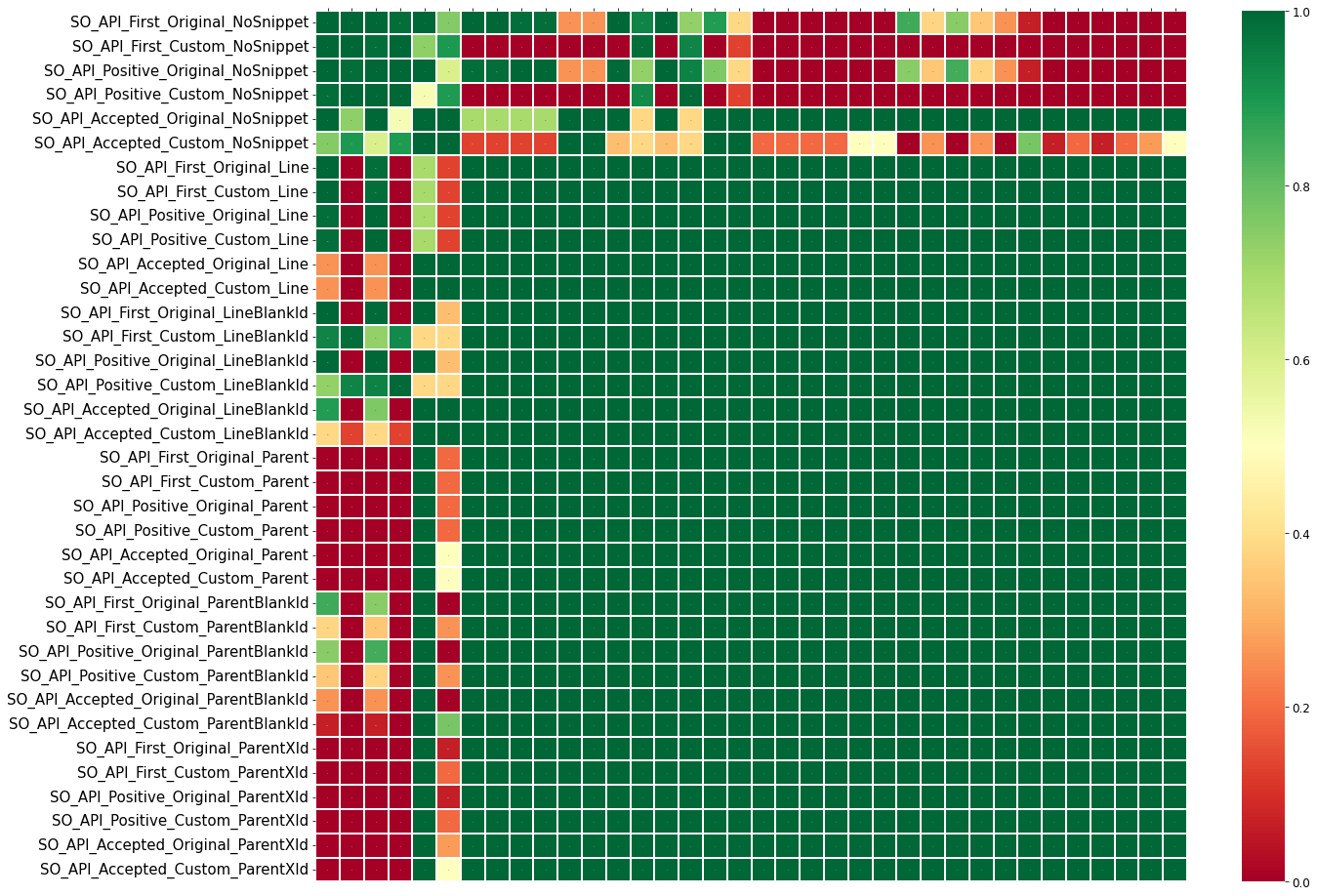}
\caption{RBO values between StackExchange \textsc{SO-API} queries}
\label{fig: rbos}
\end{figure}

Furthermore, we calculated the median RBOs between different combinations and present them in Figure~\ref{fig: rbog} and~\ref{fig: rbos}. Although we calculated all RBOs between Google and SE queries, we found the median to be 0 for all combinations with each other. This indicates their dissimilarity and provides a rationale for presenting the results in two separate figures.

\subsubsection{Impact of Query Content}

As expected, the addition of code snippets resulted in a lower median number of results. Although the median for Google queries without code snippets was more than 30,000, it was at least 10 times lower for all other combinations tested. The addition of a single line where the error was thrown reduced the median from 32,400 to 471 (-98.5\%). Including the parent node of the error line instead led to a further decrease to a median value of 6.

Although the difference was not as large, removing or replacing identifiers within code snippets with ``x'' increased the number of results compared to their ``as-is'' code snippet equivalents. Interestingly, the median number was higher when the identifiers were replaced with ``x'' (\textsc{SO-Google/ParentXId}) than when completely removed (\textsc{SO-Google/ParentBlankId}).

In terms of error message treatment, removing code specific identifiers from the error message (\textsc{SO-Google/Custom}) resulted in a higher number of results compared to simply using the original compiler error message (\textsc{SO-Google/Original}), with one exception.

Regarding relevancy, considering only the first link returned for the 100 scenarios from each combination, those returned from queries with no code snippets had the best IH-Score performance. This was observed irrespective of the query method used and whether post filtering was applied or not. The best overall results in terms of IH and IH/M-Score were achieved by both \textsc{SO-Google/First/Original/NoSnippet} and \textsc{SO-Google/Positive/Original/NoSnippet} queries at 57\% and 139\%, respectively.

For SE queries (\textsc{SO-API}), however, this came with the trade-off of a significantly higher M-Score compared to when code snippets were added. Their M-Scores averaged around 40\%, compared to other combinations, which had M-Scores between 0\% and 9\%. Consequently, this led to the IH/M-Score being less than 100\%, meaning that most of the results were irrelevant and useless for fixing the original compiler error.

For Google queries, the inclusion of code snippets had a more varied impact. One of the more evident outcomes is that adding the parent node with identifiers replaced with ``x'' as the code snippet proved to be the worst addition, with only 14\% to 22\% IH-Score and up to 62\% misleading results received from these types of queries. The addition of the unedited parent node (\textsc{SO-Google/Parent}) performed slightly better, although their IH/M-Scores were still only 48\% at best. On the other hand, when the parent node was added, but with the code specific identifiers removed entirely, it resulted in a significantly lower number of misleading results compared to all other (\textsc{SO-Google}) combinations. However, the overall IH/M-Scores remained lower than when using no code snippets.

Interestingly, this finding does not apply to \textsc{SO-API} queries, where the lowest M-Scores at 0\% occurred when the parent node was included as is or with the identifiers replaced with ``x''. Instead, removal of the identifiers increased the M-Scores to 4\% to 5\%. Overall, adding more context to SE queries led to higher IH/M-Scores, where the reduction in misleading results had a more significant impact than the relatively smaller decrease in IH results. However, this comes with the caveat that the overall number of results from SE queries, excluding those with no code snippets, is relatively low, and most of the queries return nothing.

To analyse the similarity of the results, we examine the calculated RBO values. The inclusion or exclusion of code snippets in the query, along with the type if any was included, likely affects the similarity of the results. In Figure~\ref{fig: rbog}, the presence of green and yellow $6\times6$ squares on the diagonal indicates the relatively high correlation between the first 10 results of Google queries with the same code snippets, regardless of error message treatment and post filtering. This is most apparent between queries with the parent node included as code snippets, represented by the three green squares in the bottom right quadrant.

In contrast, there are no green boxes outside the $6\times6$ area that represent those with exactly the same type of code snippet. This means that removing or replacing the identifiers within the code snippets significantly changes the results, leading to lower similarity and, consequently, RBO. However, some correlation remains visible, particularly between combinations with the error line as the code snippet (\textsc{SO-Google/Line}, \textsc{SO-Google/LineBlankId}). For these, although deleting the identifiers affected the results, the impact was less than for the parent node code snippet.

There is also a difference between removing identifiers from the error line and not doing so. A correlation can still be observed between queries with the error line without identifiers \textsc{SO-Google/LineBlankId} and a plain query without a code snippet \textsc{SO-Google/NoSnippet}, albeit only within the yellow range. In contrast, the same cannot be said between queries using the error line as is \textsc{SO-Google/Line} and those with no code snippet \textsc{SO-Google/NoSnippet} due to their low RBO with each other.

As mentioned earlier, SE queries often returned empty sets. Since the RBO value between two empty sets is 1, this resulted in a high overall RBO value and formed the large green square in Figure \ref{fig: rbos}. The small green box in the upper left corner, on the other hand, represents the RBO values between the results when querying SE without any code snippets and only the original compiler error messages \textsc{SO-API/NoSnippet}. Similarly to Google queries, the results of these queries appear to be highly correlated with each other, and the removal of identifiers from the error message has no major impact.

\subsubsection{Impact of Query Method}

We discovered that the number of results is significantly affected by the query method used. The highest median number of results for SE \textsc{SO-API} queries was only 3.5, found when searching for the generic compiler error message with the identifiers removed without any code snippets \textsc{SO-API/Custom/NoSnippet}. On the contrary, searching for the same query on Google \textsc{SO-Google/Custom/NoSnippet} yielded a median of 36,350 results. When adding any kind of code snippet, regardless of whether it was the code error line or the parent snippet, the median number of SE results was 0.

Although SE queries with code snippets rarely return anything, when they do, the results are much more likely to be accurate, with IH/M-Scores above 100\%. Queries with the parent node added as is or replaced with x (\textsc{SO-API/Parent}, \textsc{SO-API/ParentXId}) performed particularly well. Although they only returned a result four times within the 100 scenario runs, all of the first links returned were relevant.

Overall, when using SE queries with code snippets, the results are proportionally more relevant than their equivalents on Google, although the results were returned less frequently. Note that this trend does not apply to queries without code snippets \textsc{SO-API/NoSnippet}, which have IH/M-Scores below 60\%. For general searches without code snippets, Google queries had a higher percentage of relevant results, both in terms of overall numbers (IH-Score) and as a percentage over the number of misleading queries (IH/M-Score).

Taking into account the top 10 results of each query and their corresponding median RBO scores with each other, we found no similarities between the SE results and the Google results.

\subsubsection{Impact of Post Filtering}

Among the three factors discussed so far, post filtering arguably has the least noticeable impact on the relevancy of results. Specifically, for results obtained by filtering only for positive posts, no discernible impact can be observed compared to simply taking the first post returned. Interestingly, filtering for only posts with an accepted answer did not result in higher relevance; rather, it was lower in most combinations. Further analysis revealed that this occurred when the original first link found had a positively rated answer that correctly addressed the problem, but was not accepted as the answer. Consequently, the tool had to search further down the list of returned links to find a post with an accepted answer. In some cases, these posts may be less relevant to the original error, discussing a completely different problem and leading to a misleading result.

For Google results, most of the time, the first returned results are posts with positively rated answers. This is evidenced by the high RBO values between simply taking the first answer and filtering for positive answers (\textsc{SO-Google/First}, \textsc{SO-Google/Positive}) for the same query. Their RBO values range from 0.75 to 1.00, represented by the medium to dark green $4\times4$ boxes on the diagonal in Figure~\ref{fig: rbog}. Comparatively, the first results are less likely to be accepted answers, since filtering for only posts with accepted answers resulted in lower RBOs without filters. This is illustrated by the lighter green to yellow rectangles surrounding the green $4\times4$ boxes, with RBO values between 0.49 and 0.94.

\subsection*{\textbf{[RQ3]} To what extent does the query configuration influence the ability of large language models to explain compiler errors?}

Unlike the results from Google and the SE API, where sometimes the number of results returned was 0, the LLMs provided a reply for all the scenarios we tested. However, this came with the caveat that there were instances where the LLM produced ``plausible-sounding but incorrect or nonsensical answers''~\cite{openai}. In particular, there was an instance where the LLM gave the correct fix to the code, should a user choose to directly copy the recommended code fix, but provided a misleading explanation for it. Additionally, there were instances where the tool provided multiple answers, one being correct and the other misleading. Even when using the newer version, there was still a scenario where running the provided code would fix the compiler error but cause an ArrayIndexOutOfBoundsException at runtime.

Nevertheless, overall results highlighted LLM's potential, with the majority of responses received being either instrumental or helpful in fixing the queried compiler error.

As success was already seen in using LLM with full code snippets, we did not test further query alteration beyond inclusion or exclusion of whole snippets. Instead, we added a breakdown by code source to investigate whether there are particular types of code for which the LLM would work best. The results are summarised in Table~\ref{table: result rq4}.

\setcounter{subsubsection}{0}
\subsubsection{Impact of Query Content}

In all configurations and versions tested, adding the whole code snippet with compiler error as part of the query resulted in higher result relevancy. Especially in GPT-4, this led to a perfect result of 100\% instrumental or helpful. When a code snippet is not available, LLMs often revert to listing common causes of the error message, which may or may not cover the underlying cause.

\subsubsection{Impact of LLM Versions}

GPT-4 shows a notable improvement over its predecessor. This is true for all cases except where the code used was based on common errors (\textsc{custom}) and the complete code snippet was provided. In this case, GPT-3.5 had already achieved perfect performance at 100\%, leaving no room for improvement when using the newer version. For scenarios \textsc{blackbox} and \textsc{user\_study} and for the cases where no code snippets were provided, GPT-4 proved to be an advance. In particular, when the code snippet is provided (\textsc{GPT-4/Full}), understanding the context and intent of the user was demonstrated, e.g., by creating a code implementation based on commented-out parts of the code beyond fixing errors. In contrast, GPT-3.5 did not take extra context into account and provided more straightforward solutions, e.g., by not changing commented-out parts of the code.

\subsubsection{Impact of Query Phrasing}

As OpenAI acknowledged to be one of the limitations of LLMs~\cite{openai}, we observed that input phrasing affected overall results returned by the tool. Among the two tested questions, ``How can I fix the error?'' generally performed better in terms of IH-Scores. 


\begin{table*}[h]
\caption{IH count of large language model and Stack Overflow results by code source}
\label{table: result rq4}
\renewcommand{\arraystretch}{1.0} 
\setlength{\tabcolsep}{4pt} 
\newcolumntype{Y}{>{\centering\arraybackslash}X} 
\footnotesize
\begin{tabularx}{\textwidth}{p{1.5cm}|c|YY|YY|YY|YY}
\bottomrule
\hline
\rowcolor{gray!60}
\multicolumn{2}{c}{}  & \multicolumn{4}{c}{\textbf{\textsc{Original/FullSnippet}}} & \multicolumn{4}{c}{\textbf{\textsc{Original/NoSnippet}}} \\ \hline
\rowcolor{gray!30}
\multicolumn{2}{c}{}  & \multicolumn{2}{c}{\textbf{GPT-4}} & \multicolumn{2}{c}{\textbf{GPT-3.5}} & \multicolumn{2}{c}{\textbf{GPT-4}} & \multicolumn{2}{c}{\textbf{GPT-3.5}} \\ \hline
\rowcolor{gray!10}
\textbf{Source} & \textbf{Ttl.} & \textbf{\textsc{What}} & \textbf{\textsc{How}} & \textbf{\textsc{What}} & \textbf{\textsc{How}} & \textbf{\textsc{What}} & \textbf{\textsc{How}} & \textbf{\textsc{What}} & \textbf{\textsc{How}} \\ \hline
\textsc{blackbox} & 55 & 55 (100\%) & 55 (100\%) & 47 (85\%) & 51 (93\%) & 44 (80\%) & 46 (84\%) & 43 (78\%) & 44 (80\%) \\
\textsc{custom} & 18 & 18 (100\%) & 18 (100\%) & 18 (100\%) & 18 (100\%) & 16 (89\%) & 16 (89\%) & 13 (72\%) & 15 (83\%) \\
\textsc{user\_study} & 27 & 27 (100\%) & 27 (100\%) & 22 (81\%) & 21 (78\%) & 22 (81\%) & 22 (81\%) & 16 (59\%) & 16 (59\%) \\ \hline
\rowcolor{gray!10}
 & \textbf{100} & \textbf{100 (100\%)} & \textbf{100 (100\%)} & \textbf{87 (87\%)} & \textbf{90 (90\%)} & \textbf{82 (82\%)} & \textbf{84 (84\%)} & \textbf{72 (72\%)} & \textbf{75 (75\%)} \\ \bottomrule
\end{tabularx}
\end{table*}

\section{Implications}
\label{sec:implications}

\subsection{For Developers}

Developers spend a significant amount of time searching the Web for help and reformulating queries~\cite{r3, r5, r13}. Through our findings, we aim to improve this experience and increase overall efficiency when using LLMs or SO to find help fixing compiler errors.

When using SO, we find that searching SO via Google instead of a direct search impacts the type and number of results returned. When no code snippet is available, a Google search is considered a better approach. For queries that include only the compiler error message and \textsc{NoSnippet}, we found that Google queries had a higher percentage of helpful answers and a lower percentage of misleading results. Conversely, if a user has access to the original code, it may prove beneficial to search directly on SO while including a code snippet. We identified a notably higher percentage of helpful against misleading results for these types of queries, making this approach particularly useful for novices with less experience in gauging the relevancy and correctness of a post. However, we also note that \textsc{SO-API} rarely returns a result, so a Google search may still be needed in most cases. Moreover, removing code-specific identifiers and/or replacing them with ``x'' was often seen to increase the number of results.

When examining the first link returned from queries, we did not observe a positive relationship between filtering for a post with an accepted answer and a more relevant answer. Therefore, we suggest not discounting posts just because they contain no accepted answers.

LLMs demonstrated great potential to help developers fix their compiler errors. When comparing results across types of source code, LLMs appear to work particularly well for shorter code snippets with common errors, given the perfect results found for \textsc{custom} code snippets, and also when the entire code snippet is made available as part of the query. Thus, for developers who create simple code snippets that need a quick fix, an LLM may be an excellent tool to try first, ensuring that the code is included when available. However, unlike SO, LLM answers do not have other users voting or providing comments on accuracy. This, combined with the fact that LLMs can make misleading answers sound plausible, means that users must exercise discretion and consider whether the answers provided are truly correct.

For tool developers working with compiler errors, many adopt the approach of imitating how programmers search the Web for solutions and then automating the process. For these types of approaches, the same recommendations apply. For instance, if a tool functions as an IDE plugin, it might be beneficial to first try including code snippets in the query and searching directly on SO. On the contrary, for tools that operate similarly to a search engine and do not have access to user code, implementing a search that works via Google may be more effective.

However, this might not necessarily hold true for all approaches, and hence our findings can be implemented depending on how a tool developer plans to interact with SO and user code. It is essential for a tool developer to test different query formats and methods before committing to one, due to the dissimilarity of the results, as shown in Figure~\ref{fig: rbog} and~\ref{fig: rbos}.

As we did not test the ChatGPT API, we cannot comment on the differences compared to using it through the user interface. However, assuming similar performance, the ChatGPT API may be a viable option for developers, especially if access to the underlying code is available.

\subsection{For Researchers}

There are opportunities to further investigate the impact of different factors on query effectiveness for both SO and LLMs. Given that minor changes can have a significant impact on results, this opens up possibilities for researchers to delve deeper into the topic, e.g., by testing various parts of code snippets and/or code pre-processing approaches. Unlike Monperrus and Maia's study~\cite{r17}, where both code pre-processing methods improved their overall debugging process, we found mixed results. Consequently, further research would be beneficial, particularly considering the high percentage of SO usage in debugging errors.

Additionally, our study has demonstrated LLMs' potential in assisting with fixing compiler errors. Further research in this area would be compelling and can be approached from various angles, such as focusing on other types of errors, identifying the best types of questions to ask, or even detecting the characteristics that distinguish relevant and correct answers from misleading and non-sensical ones.

\section{Threats to Validity}
\label{sec:threats to validity}

We acknowledge that there exist limitations and threats that may impact the validity of our results.

\begin{itemize}
\item\textbf{Construct Validity:} A threat comes from the subjectivity of measuring the relevance of SO posts and LLM responses to the underlying compiler error. To mitigate this, we implemented a clearly defined set of classifications consistent with previous studies~\cite{r20, o2,o20,o15}. To further build confidence in the ratings, we also had a subset of the results independently annotated by a different rater. The inter-rater agreement of 0.860 suggests ``almost perfect'' agreement. 

\item\textbf{Only first links are considered:} When analysing the scores and metrics used in the SO results and discussion (excluding RBOs), we evaluated the results based on the first links returned from each query. One could argue that the relevance of a query should consider more than just the first link, noting that experienced developers are likely to consider further links~\cite{r1}. We recognise this as a limitation of our current approach and that our results are based on the assumption that the first link returned is an accurate representation of the relevancy of subsequent links.

\item\textbf{Only first error encountered are considered:} For the Java code from which we collected the data, queries were created only based on the first error encountered, even when their compilation leads to multiple errors. It has often been advised to fix the first error first before recompiling, and that many subsequent errors are caused by the first~\cite{t25, t26}. Hence, fixing only the first error each compile likely mimics common behaviour when error fixing anyway, and we do not anticipate this threat to have a large impact on overall validity. However, aspects of fixing multiple errors at once, a behaviour shown a few times in the LLM responses we analysed, could be a potential area for future study.

\item\textbf{Limited Number of Results from SO-API Searches:} The low number of results returned from SO-API searches can potentially impact the findings. However, developers often access Stack Overflow by querying search engines such as Google instead of directly visiting the website ~\cite{r10}. Considering that Google searches may better reflect actual search behavior, this limitation may be mitigated to some extent.

\item\textbf{Reproducibility:} As Google’s search algorithm is not public, the exact same query may return different results depending on location, time, and device type~\cite{google}, among others. This is also true for LLM answers, which are ``sensitive to tweaks to the input phrasing or attempting the same prompt multiple times''~\cite{openai}. To minimise the potential bias introduced by prompting, simple prompts were used and alternatives were evaluated. Additionally, the consistency of the methodology and the environment was maintained during data collection.

\item\textbf{Generalisability:} Our study focuses on Java compiler errors. This means that the same conclusions may not hold for other programming languages and types of errors. Additionally, even for Java compiler errors, we cannot claim that the findings will generalise beyond our dataset. Although we have tried to minimise this threat by including code from different sources, we acknowledge that the scale of the dataset itself is still small, covers relatively shorter code snippets, and may not cover all types of compiler errors. Lastly, although we observe trends, e.g., between the usefulness of crowd-curated compiler error help and automatically generated support, we cannot generalise to future versions of tools.
\end{itemize}

\section{Conclusion}
\label{sec:conclusion}

Compiler error messages often fail to provide enough targeted information to enable programmers to fix their code. Our study systematically compares SO and LLMs, such as GPT-3.5 and GPT-4, for their effectiveness in explaining compiler error messages. Our findings demonstrate that GPT-4 consistently outperforms both GPT-3.5 and SO, especially when provided with the offending code snippet along with the error message. Furthermore, we examine the influence of query configuration on SO's ability to yield relevant answers and the impact of prompt configuration on LLMs' capacity to explain compiler errors. The insights gained from this research have practical implications for programmers, who can leverage these findings to optimise their debugging process, and for researchers, who can build upon this work to explore the rapidly evolving landscape of AI-driven assistance in programming and debugging.

\bibliography{addressing_compiler_errors}





\end{document}